# Unconventional Hall Effect induced by Berry Curvature

Jun Ge[1,†], Da Ma[1,†], Yanzhao Liu[1], Huichao Wang[1,2], Yanan Li[1,3], Jiawei Luo[1], Tianchuang Luo[1], Ying Xing[1,4], Jiaqiang Yan[5,6], David Mandrus[5,6], Haiwen Liu[7], X.C. Xie[1,8,9,10,*], Jian Wang[1, 8,9,10, *]

[1] *International Center for Quantum Materials, School of Physics, Peking University, Beijing 100871, China*
[2] *Department of Applied Physics, The Hong Kong Polytechnic University, Kowloon, Hong Kong, China*
[3] *Department of Physics, Pennsylvania State University, University Park, PA 16802*
[4] *Department of Materials Science and Engineering, School of New Energy and Materials, China University of Petroleum, Beijing 102249, China*
[5] *Department of Materials Science and Engineering, University of Tennessee, Knoxville, Tennessee 37996, USA*
[6] *Materials Science and Technology Division, Oak Ridge National Laboratory, Oak Ridge, Tennessee 37831, USA*
[7] *Center for Advanced Quantum Studies, Department of Physics, Beijing Normal University, Beijing 100875, China*
[8] *CAS Center for Excellence in Topological Quantum Computation, University of Chinese Academy of Sciences, Beijing 100190, China*
[9] *Beijing Academy of Quantum Information Sciences, Beijing 100193, China*
[10] *Collaborative Innovation Center of Quantum Matter, Beijing 100871, China*
[†]These authors contributed equally to this work.
*email: jianwangphysics@pku.edu.cn (J.W.); xcxie@pku.edu.cn (X.C.X.)

**Berry phase and Berry curvature play a key role in the development of topology in physics and do contribute to the transport properties in solid state systems. In this paper, we report the finding of novel nonzero Hall effect in topological material $ZrTe_5$ flakes when in-plane magnetic field is parallel and perpendicular to the current. Surprisingly, both symmetric and antisymmetric components with respect to magnetic field are detected in the in-plane Hall resistivity. Further theoretical analysis suggests that the magnetotransport properties originate from**

**the anomalous velocity induced by Berry curvature in a tilted Weyl semimetal. Our work not only enriches the Hall family but also provides new insights into the Berry phase effect in topological materials.**

**INTRODUCTION**

The concept of Berry phase was firstly proposed in 1984 and has led to significant breakthroughs in physical science [1]. Recent years, Berry curvature, first discussed in the one band effective dynamics of a Bloch electron [2, 3], has become an important concept in condensed matter physics [4-13]. Numerous experimental investigations show that it has to be considered in the semiclassical electronic theory as a basic ingredient. Many interesting physical phenomena can be generated by Berry curvature, such as the intrinsic anomalous Hall effect [14-18], negative magnetoresistance (NMR) [19, 20] and nonlinear Hall effect [21, 22]. In time-reversal symmetry broken or space-inversion symmetry broken systems, the nonzero Berry curvature can generate an anomalous velocity that is transverse to the applied electric field and thus gives rise to anomalous transport currents, making intrinsic contribution to Hall conductivity [2, 4]. Besides, Berry curvature can also affect the density of states in the phase-space, which leads to the violation of Liouville's theorem for the conservation of phase-space volume [23]. As a consequence, a correction term to the density of states dependent on magnetic field and Berry curvature is generated, which has profound effects on transport properties.

The emergence of Weyl semimetal provides a new platform to explore Berry curvature. Weyl semimetals host gapless bulk excitations described by Weyl equation and metallic Fermi arc states on the surface. As a result of band crossing between a pair of spin-non-degenerate bands, Weyl node behaves like magnetic monopole in momentum space and can generate divergent Berry curvature around the Weyl points [24-26]. The Berry curvature integrating over the Fermi surface enclosing a Weyl point gives a quantized topological charge, which equals the chirality of Weyl node [27]. Many interesting Berry curvature-related physical phenomena are proposed in Weyl semimetals e.g. chiral anomaly [27, 28]. Recently, it was proposed that the Berry curvature can give rise to a new type planar Hall effect (PHE) in Weyl semimetals [29, 30]. The proposed PHE appears when the electric and magnetic field are coplanar, symmetric with magnetic field, and satisfies the angular relation:

$\sigma_{yx}^{ph} = \Delta\sigma sin\theta cos\theta$, with $\theta$ denoting the angle between electric and magnetic field. When coplanar electric and magnetic field are aligned in parallel and perpendicular directions, the proposed PHE signal will vanish. The systematic experimental studies on the relation between the PHE and the magnetic field, especially on the symmetric-antisymmetric property and the angular dependence, are still highly desired.

In this work, we present the first experimental observation of nonzero Hall effect when in-plane magnetic field is parallel and perpendicular to the current, which is clearly revealed by systematic and reliable magnetotransport studies in topological material ZrTe$_5$ devices. Specifically, when magnetic field lies in *ac* plane of ZrTe$_5$ flakes, both symmetric and antisymmetric components are observed in Hall resistivity, which cannot be explained by previous consideration of the PHE. When in-plane magnetic field is along *a* axis ($B \parallel I$) and *c* axis ($B \perp I$), nonzero in-plane Hall resistivity is detected after excluding extrinsic contributions from longitudinal magnetoresistance caused by slight misalignment of electrodes and inevitable projection of normal Hall resistivity at *b* axis. This feature is also beyond the current theoretical understanding and experimental reports on the PHE, for which the Hall signal is zero when the coplanar magnetic field and current are parallel and perpendicular [29-34]. Our theoretical analysis reveals that the tilt term in Weyl semimetal gives rise to extra contribution to the Berry curvature-induced anomalous velocity, which can explain both the symmetric-antisymmetric properties and the angular dependence of Hall signals in our measurements. Thus, our findings reveal the new properties induced by Berry curvature.

**RESULTS**

ZrTe$_5$ is a layered material, with two-dimensional sheets in *ac* plane stacking along the *b* axis, and the interlayer van der Waals coupling strength is similar to that of graphite [35]. Flat ZrTe$_5$ flakes can thus easily be obtained by mechanical exfoliation.

Many experimental investigations have indicated that ZrTe$_5$ can be a three-dimensional Dirac semimetal [36-39] depending on the unit cell volume of the specific crystals, and lots of interesting phenomena induced by nontrivial Berry curvature have been observed in ZrTe$_5$, such as chiral magnetic effect [36], recently reported PHE [31] and anomalous Hall effect [40]. In this work, ZrTe$_5$ flakes supported by 300 nm-thick SiO$_2$/Si substrates were mechanically exfoliated from high quality single crystals [38]. The standard e-beam lithography followed by e-beam evaporation was used to fabricate electrodes. Figure 1a shows the atomic force microscope (AFM) image of a typical device for electrical transport measurements. The crystal orientation is indicated by the white arrows. ZrTe$_5$ flakes are often narrow in the $c$ axis as they are easily peeled off along $a$ axis. Refined Hall structure configuration was used to detect Hall signals of ZrTe$_5$ devices on $ac$ plane. Electric current is always along $a$ axis and homogeneously goes through the device as current electrodes ($I_+$ and $I_-$) cover the whole flake. The typical longitudinal resistivity as a function of temperature from 300 K to 2 K is shown in Fig. 1(b). Device 1(s1, 300 nm thick) and device 2(s2, 225 nm thick) are selected as representatives here. With temperature decreasing from 300 K to 2 K, a resistivity peak can be clearly observed in both devices. This resistivity vs. temperature ($\rho T$) behavior is different from the resistivity saturation at low temperatures for the bulk single crystals [38], from which the devices are exfoliated. Another feature shown in $\rho T$ behavior is that the thinner flake exhibits higher temperature of the resistivity peak, which indicates the enhancement of the metallic state in thinner flakes. Previous studies show that this is a consequence of the energy bands shifting [41].

We further explore the magnetotransport properties of ZrTe$_5$ flakes in the measurement configuration when magnetic field is aligned in $ac$ plane. Figure 2(a) shows the schematic view of the angular-dependent magnetotransport measurement configuration. The in-plane magnetic field lies in $ac$ plane and forms an angle $\theta$ relative to the electric current. Magnetic field is along $a$ axis for $\theta = 0°$ ($B \parallel I$) and along $c$ axis for $\theta = 90°$ ($B \perp I$). Figure 2(b) exhibits the longitudinal magnetoresistance (LMR) at selected angles taken in s2. Negative longitudinal magnetoresistance is most evident at $a$ axis ($B \parallel I$), which is consistent with the prediction of chiral anomaly. Moreover, negative LMR is detected in a large regime of about $\pm 20°$, several times larger than that reported in bulk ZrTe$_5$ [36, 40]. It may be due to the more precise angle control for ZrTe$_5$ thin flakes. Figure 2(c) shows the in-plane Hall resistivity after excluding contribution from LMR caused by slight misalignment of electrodes, and the formula of the renormalized Hall resistivity reads $\rho_{yx} = \rho_{yx}^{raw}(B) - \rho_{xx}(B) \frac{\rho_{yx}^{raw}(0)}{\rho_{xx}(0)}$. The raw Hall resistivity at selected angles can be found in Fig. S1. As shown in Fig. 2(c), the Hall traces exhibit both symmetric and antisymmetric components. More interestingly, nonzero Hall resistivity at both $B \parallel I$ and $B \perp I$ are clearly detected. This is forbidden in classical theory. The Hall resistivity should vanish at $\theta = 0°(B \parallel I)$ and $\theta = 90°(B \perp I)$ even considering the current model for planar Hall resistivity in Weyl semimetals [29, 30].

To understand the detected in-plane Hall signals, we separately explore the symmetric and antisymmetric components in Hall traces. The symmetric component of the in-plane Hall resistivity is obtained by the symmetrization $\rho^s{}_{yx} = \frac{\rho_{yx}(+B) + \rho_{yx}(-B)}{2}$. Figure 3(a) displays the symmetric in-plane Hall resistivity $\rho^s{}_{yx}(B)$ at selected angles in s2. As the magnetic field is increased from 0 T to 6 T, the symmetric in-plane Hall resistivity grows and finally saturates at large magnetic field at most angles. More importantly, nonzero symmetric in-plane Hall resistivity at $B \parallel I$ ($\theta = 0°$) and $B \perp I$ ($\theta = 90°$) axis can be clearly observed. The temperature dependence of the symmetric in-plane Hall resistivity $\rho^s{}_{yx}(B)$ at $B \parallel I$ and $B \perp I$ is shown in Fig. 3(b) and Fig. 3(c), respectively. At a fixed magnetic field, $\rho^s{}_{yx}(B)$ at both $B \parallel I$ and $B \perp I$ decreases as temperature increases, and finally vanishes at about 200 K. To study how in-plane Hall behavior changes with the orientation between magnetic field and current, we detect the symmetric in-plane Hall resistivity as a function of $\theta$ at various magnetic fields. Figure 3(d) illustrates the symmetric in-plane Hall resistivity taken at various magnetic fields. Same symmetrization process is carried out to remove the contribution from antisymmetric component. At $\theta = 0°(B \parallel I)$ and $\theta = 90°(B \perp I)$, the nonzero symmetric in-plane Hall resistivity can be clearly detected, which is in good consistence with the Hall results in Fig. 3a.

To obtain intrinsic antisymmetric in-plane Hall resistivity, we carried out angular-dependent magnetotransport measurements in *ac* plane of ZrTe$_5$ device s2 in a triple axes vector magnet. Figure 4(a) gives the antisymmetric in-plane Hall resistivity at selected angles taken at 1.1 K. Magnetic field lies in *ac* plane and is tilted away from the *a* axis ($\theta = 0°$). We define antisymmetric in-plane Hall resistivity as $\rho^{as}_{yx} = \frac{\rho_{yx}(+B) - \rho_{yx}(-B)}{2}$ to remove the contribution from symmetric component. At low magnetic field region between -0.7 T and 0.7 T, linear behavior is observed at all angles. Nonzero antisymmetric in-plane Hall resistivity is detected at both $\theta = 0°(B \parallel I)$ and $\theta = 90°(B \perp I)$. We solely display the antisymmetric in-plane Hall resistivity near $B \parallel I$ and $B \perp I$ in Fig. 4(b) and Fig. 4(c) for clarity. As a comparison, Fig. S3 shows the antisymmetric in-plane Hall resistivity measured in PPMS after the same antisymmetrization process. Hall resistivity in the same magnetic field region from -0.7 T to 0.7 T is exhibited in Fig. S3(b), which is consistent with data in Fig. 3. Figure 4(d) and Fig. 4(e) show temperature dependence of antisymmetric in-plane Hall resistivity $\rho^{as}_{yx}(B)$ at $B \parallel I$ and $B \perp I$. A linear $\rho^{as}_{yx}(B)$ behavior is clearly obtained. In low temperature regime, Hall resistivity at fixed magnetic fields slightly decreases with increasing temperature at both $B \parallel I$ and $B \perp I$.

To further confirm the detected nonzero Hall signals when in-plane magnetic field is parallel and perpendicular to the current, we carry out Hall measurements in other ZrTe$_5$ devices too. As shown in Fig. S5, similar nonzero Hall resistivity can be observed in device s1. Therefore, in topological ZrTe$_5$ devices, Hall effect does exist when in-plane magnetic field is parallel and perpendicular to the current.

The nonzero Hall resistivity detected at $B \parallel I$ and $B \perp I$ is very unusual. In fact, this unconventional behavior can be understood as the PHE in a topological semimetal. Previous studies on the PHE show that the conductivities depend quadratically on the magnetic field, and the Hall signal is zero when the electric field and the magnetic field are parallel and perpendicular [29, 30]. However, we find that there are linear conductivities and nonzero Hall signal at $B \parallel I$ and $B \perp I$ in the PHE of tilted Weyl semimetals, where the anomalous velocity, the chiral chemical potential, and the phase volume factor all play a part. ZrTe$_5$ is considered as a topological semimetal in magnetic field [38]. Specifically, we investigate a low energy effective model [37] based on previous density functional theory calculation [35] and find that magnetic field drives ZrTe$_5$ into a tilted Weyl semimetal phase with a pair of Weyl points. (See the Supplementary Information for details.) This is not surprising, since tilt is common among Weyl semimetals, as Weyl points are typically low-symmetry points [42, 43]. We perform a semiclassical calculation on the conductivities of a tilted Weyl semimetal and compare the theoretical findings with the experimental results. (See the Supplementary Information for details.) As shown in Fig. 4(f), the black circle data points are the angular dependence of observed in-plane Hall conductivity detected in vector magnet at 1.1 K and 1 T after excluding $\rho_{xx}$. The red theoretical fitting curve with a set of reasonable parameters (listed in Supplementary Information) is well consistent with the experimental data [39].

The Dirac cones in ZrTe$_5$ may open a small gap $\Delta$ of the order several meV in *ab initio* calculations [26] considering the precise value of the spin orbital coupling strength. Nevertheless, in our samples $E_F \gg \Delta$ is satisfied, indicating that the gap has little impact on the Berry curvature and the Fermi velocity at the Fermi surface. Therefore, our theoretical conclusion stands even considering a small gap in ZrTe$_5$.

**CONCLUSIONS**

In summary, we fabricated and measured ZrTe$_5$ devices with standard Hall bar structure. The detected Hall signals exhibit both symmetric and antisymmetric behavior when the magnetic field is in *ac* plane. After removing the contributions from LMR caused by slight misalignment of electrodes and projection of normal Hall with $B \parallel b$ axis, nonzero in-plane Hall resistivity is clearly obtained when in-plane magnetic field is parallel and perpendicular to the current, which is unprecedented in previous theoretical and experimental studies. Theoretical calculation suggests that the unconventional behavior of the Hall conductivity can be understood within the semiclassical transport of the tilted Weyl semimetals, where the chiral chemical potential, the Berry curvature-induced anomalous velocity and the phase space volume correction all take part. Our discovery of nonzero in-plane Hall signals at $B \parallel I$ and $B \perp I$ not only adds a new member to Hall effect family but also paves a new way to explore Berry phase and Berry curvature in topological materials.

**METHODS**

**Fabrication of ZrTe$_5$ flake devices**. The ZrTe$_5$ bulk crystals were grown as described in ref.38. ZrTe$_5$ flakes supported by 300 nm-thick SiO$_2$/Si substrates were mechanically exfoliated from high quality single crystals with scotch tape. The detailed fabrication process of electrodes is described below.

1. Standard Electron beam lithography in a FEI Helios NanoLab 600i DualBeam System to define electrodes after spin-coating PMMA resist.
2. Develop resist in a IPA: MIBK (3:1 by weight) mixture.
3. Deposit Pd (6.5 nm)/Au (300 nm) in a LJUHV E-400L E-Beam Evaporator after Ar plasma cleaning.
4. Liftoff with acetone.

**Transport measurements**. Electrical transport measurements are conducted in a 16T-Physical Property Measurement System (PPMS-16T) from Quantum Design at 2 K and a Leiden dilution refrigerator (CF450) with a triple axes vector magnet at 1.1 K.

**FUNDING**

This work was financially supported by the National Key Research and Development Program of China (2018YFA0305604 and 2017YFA0303300), the National Natural Science Foundation of China (Grant No. 11888101 and Grant No. 11774008), the Strategic Priority Research Program of Chinese Academy of Sciences (XDB28000000), and Beijing Natural Science Foundation (Z180010).

**AUTHOR CONTRIBUTIONS**

J. W. conceived the research and supervised the experiments. J.G., Y.Liu, H.W., Y.Li, J.L., T. L. performed the electrical transport measurements and analyzed the data. D.Ma, H.L., and X.C.X developed the theoretical model. J.Y. and D.Mandrus grew the bulk crystals. J.G. and Y.Li. fabricated the devices. All authors offered the discussion or revisions on the manuscript.

**CONFLICT OF INTEREST**

The authors declare that they have no conflict of interest.

Figures

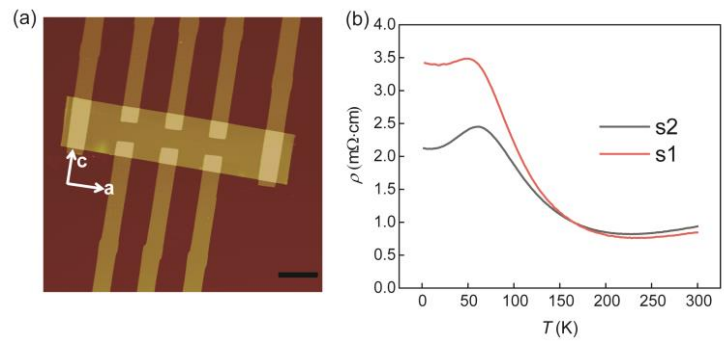

**Figure 1.** AFM image and $\rho$(T) behavior of ZrTe$_5$ device. (a) AFM image of a typical ZrTe$_5$ device. The scale bar represents 10 μm. (b) Resistivity of ZrTe$_5$ devices (s1:300 nm thick; s2:225 nm thick) as a function of temperature from 300 K to 2 K. A resistivity peak is detected.

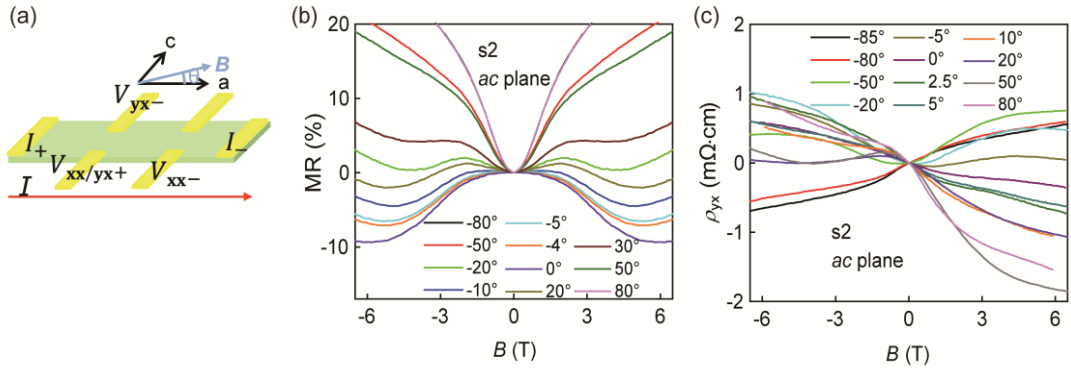

**Figure 2.** Angular dependence of magnetoresistance ratio and in-plane Hall resistivity in s2. (a) Schematic structure for the angular-dependent magnetotransport measurements in *ac* plane. In-plane magnetic field is along *a* axis for $\theta = 0°$ ($B \parallel I$) and along c axis for $\theta = 90°$ ($B \perp I$). (b) MR behavior at selected angles. Negative LMR is detected in a large angular regime of $\pm 20°$ and reaches maximum at $\theta = 0°$ ($B \parallel I$). (c) in-plane Hall data after subtracting $\rho_{xx}$ caused by slight electrodes misalignment through $\rho_{yx} = \rho_{yx}^{raw}(B) - \rho_{xx}(B) \cdot \frac{\rho_{yx}^{raw}(0)}{\rho_{xx}(0)}$. Both symmetric and antisymmetric contribution can be observed in $\rho_{xy}$.

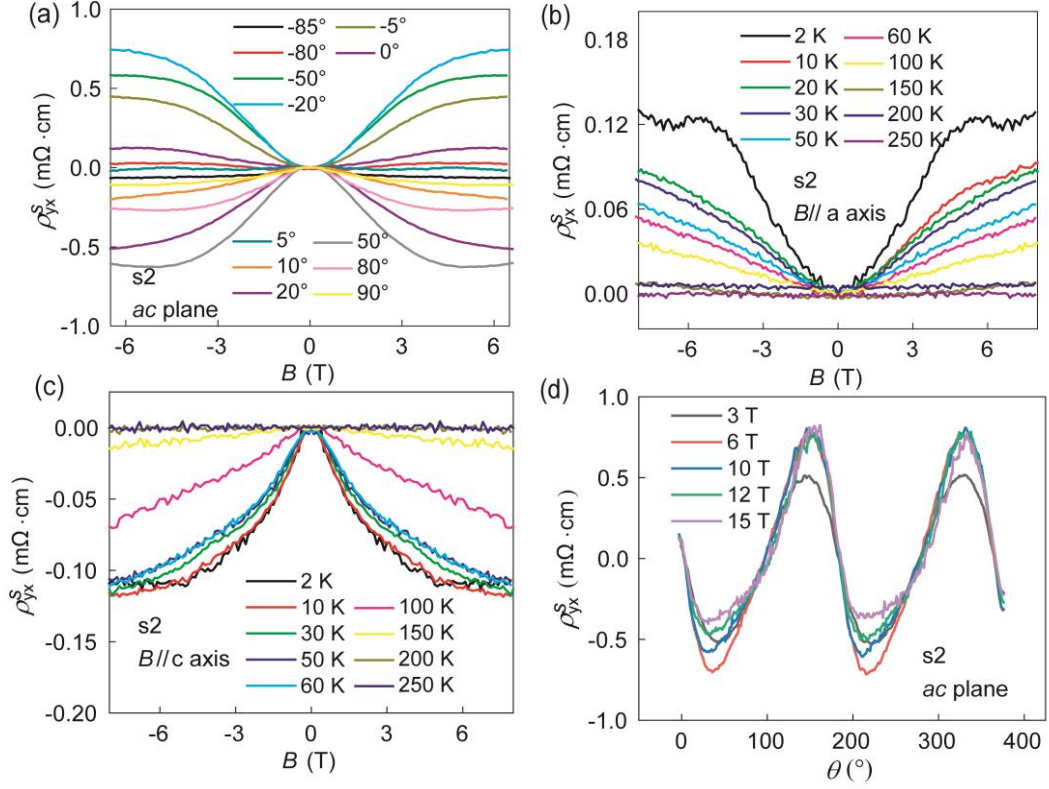

**Figure 3.** In-plane Hall signals detected in ZrTe$_5$ device s2 after symmetrization. Magnetic field lies in *ac* plane and is tilted away from the *a* axis ($\theta = 0°$). (a) Symmetric in-plane Hall resistivity vs. *B* at selected angles. Symmetric in-plane Hall resistivity grows as magnetic field increase, and saturates at high magnetic field. (b),(c) Symmetric in-plane Hall resistivity vs. *B* at $B \parallel I$ ($\theta = 0°$) and $B \perp I$ ($\theta = 90°$) at various temperatures. At fixed magnetic field, symmetric in-plane Hall resistivity decreases as temperature increases, finally vanishes at about 200 K. (d) Symmetric in-plane Hall resistivity as a function of $\theta$ at different magnetic fields (T=2 K) with $\rho^s_{yx} = \frac{\rho_{yx}(+B)+\rho_{yx}(-B)}{2}$. All curves indicate nonzero planar Hall resistivity at $B \parallel I$ ($\theta = 0°$) and $B \perp I$ ($\theta = 90°$).

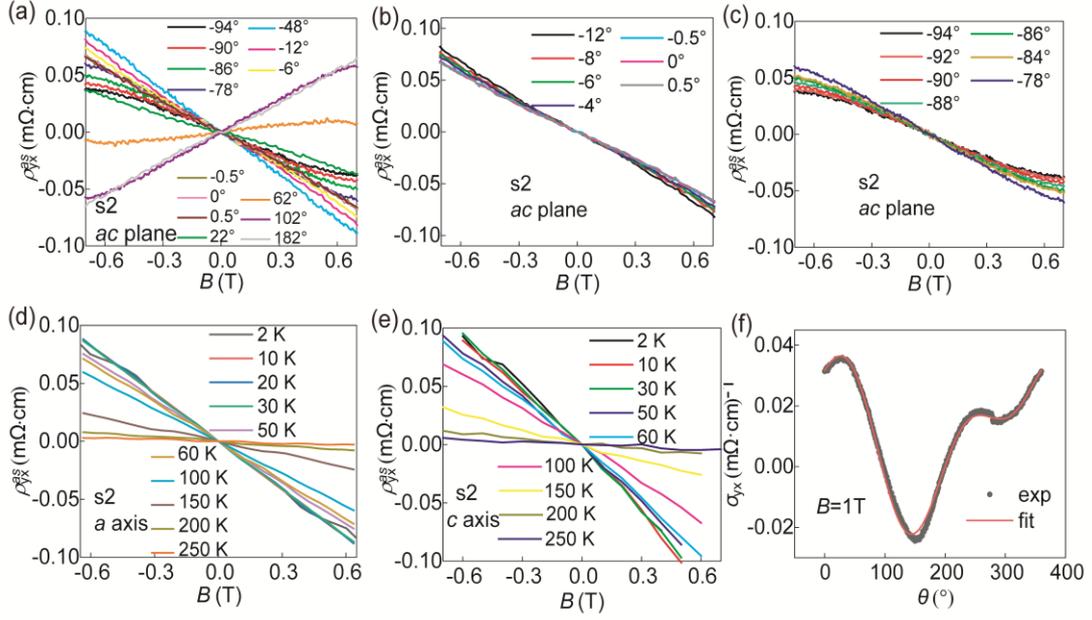

**Figure 4.** In-plane Hall resistivity detected in ZrTe$_5$ device s2 after antisymmetrization and theoretical fitting of the observed in-plane Hall conductivity. Magnetic field lies in *ac* plane and is tilted away from the *a* axis($\theta = 0°$). (a) Antisymmetric in-plane Hall resistivity vs. *B* at selected angles. Linear behavior at low magnetic field can be observed. (b),(c) Antisymmetric in-plane Hall resistivity versus *B* around $B \parallel I$ ($\theta = 0°$) and $B \perp I$ ($\theta = 90°$). Antisymmetric in-plane Hall signal is also apparent at both $B \parallel I$ ($\theta = 0°$) and $B \perp I$ ($\theta = 90°$). (d),(e) Antisymmetric in-plane Hall resistivity versus *B* at $B \parallel I$ ($\theta = 0°$) and $B \perp I$ ($\theta = 90°$) at various temperatures. All Hall traces show linear behavior in low field region. (f) Theoretical fitting (red curve) of the in-plane Hall conductivity.

Supplementary Information for

# Unconventional Hall Effect induced by Berry Curvature

Jun Ge[1,†], Da Ma[1,†], Yanzhao Liu[1], Huichao Wang[1,2], Yanan Li[1,3], Jiawei Luo[1], Tianchuang Luo[1], Ying Xing[1,4], Jiaqiang Yan[5,6], David Mandrus[5,6], Haiwen Liu[7], X.C. Xie[1,8,9,10,*], Jian Wang[1, 8,9,10, *]

[1] *International Center for Quantum Materials, School of Physics, Peking University, Beijing 100871, China*
[2] *Department of Applied Physics, The Hong Kong Polytechnic University, Kowloon, Hong Kong, China*
[3] *Department of Physics, Pennsylvania State University, University Park, PA 16802*
[4] *Department of Materials Science and Engineering, School of New Energy and Materials, China University of Petroleum, Beijing 102249, China*
[5] *Department of Materials Science and Engineering, University of Tennessee, Knoxville, Tennessee 37996, USA*
[6] *Materials Science and Technology Division, Oak Ridge National Laboratory, Oak Ridge, Tennessee 37831, USA*
[7] *Center for Advanced Quantum Studies, Department of Physics, Beijing Normal University, Beijing 100875, China*
[8] *CAS Center for Excellence in Topological Quantum Computation, University of Chinese Academy of Sciences, Beijing 100190, China*
[9] *Beijing Academy of Quantum Information Sciences, Beijing 100193, China*
[10] *Collaborative Innovation Center of Quantum Matter, Beijing 100871, China*
[9] *Beijing Academy of Quantum Information Sciences, Beijing 100193, China*
[10] *Collaborative Innovation Center of Quantum Matter, Beijing 100871, China*

[†]These authors contributed equally to this work.
*email: jianwangphysics@pku.edu.cn (J.W.); xcxie@pku.edu.cn (X.C.X.)

## Supplementary notes on theoretical details

### I. Semiclassical transport and theoretical fitting

We investigate the electronic transport in ZrTe$_5$ in the semiclassical regime. With Berry curvature $\Omega_k$ included, the equations of motion read [1-4]

$$\dot{r} = \frac{1}{\hbar}\frac{\partial \epsilon}{\partial k} - \dot{k} \times \Omega_k, \#(S1)$$

$$\hbar\dot{k} = -eE - e\dot{r} \times B, \#(S2)$$

where $\dot{r}$ is the group velocity of the wave packet, $k$ is the momentum, $E$ is the electric field, and $B$ is the magnetic field. From the equations above, we can obtain

$$\dot{r} = D(B,\Omega_k)[v_k + \frac{e}{\hbar} E \times \Omega_k + \frac{e}{\hbar}(v_k \cdot \Omega_k)B], \#(S3)$$

$$\hbar\dot{k} = -D(B,\Omega_k)[eE + ev_k \times B + \frac{e^2}{\hbar}(E \cdot B)\Omega_k], \#(S4)$$

where $v_k = \frac{1}{\hbar}\frac{\partial \epsilon}{\partial k}$ is the Fermi velocity, and $D(B,\Omega_k) = \left[1 + \frac{e}{\hbar}(B \cdot \Omega_k)\right]^{-1}$ is the phase-space volume factor [1, 4]. In the presence of a magnetic field $B$ and a Berry curvature $\Omega_k$, the phase space volume is not conserved, and the current density becomes [1, 4]

$$J = -e\int \frac{dk}{(2\pi)^3} D^{-1}\dot{r} f_k(r), \#(S5)$$

where $f_k(r)$ is the distribution function of the electrons occupying the $k$ state at the location $r$ in real space, which is calculated from the Boltzmann equation with a relaxation time approximation,

$$\dot{k} \cdot \nabla_k f_k(r) = \frac{f_{eq}(r) - f_k(r)}{\tau(k)}. \#(S6)$$

Here $\tau(k)$ is the relaxation time, and $f_{eq}(r)$ is the distribution function at equilibrium. Substituting Eq. (S4) into Eq. (S6), we find

$$f_k = f_{eq} + \left[eD\tau E \cdot v_k + \frac{e^2}{\hbar} D\tau(E \cdot B)(v_k \cdot \Omega_k) + v_k \cdot \Gamma\right]\frac{\partial f_{eq}}{\partial \epsilon}, \#(S7)$$

where $\Gamma$ is the higher-order correction term.

Next, we consider the Weyl semimetal with tilt term. For a pair of Weyl cones related by symmetry, the Hamiltonian around the Weyl point can be written as

$$H = \hbar v k \cdot \sigma + \hbar t \cdot k \#(S8)$$

for one cone, and

$$H = -\hbar v \boldsymbol{k} \cdot \boldsymbol{\sigma} - \hbar \boldsymbol{t} \cdot \boldsymbol{k} \#(S9)$$

for the other cone with the opposite chirality. Here $\boldsymbol{\sigma}$ is the vector of the Pauli matrices, and $\boldsymbol{t}$ is the tilt vector. Here we suppose that the tilt lies in the $xy$ plane, and thus the tilt term reads

$$\boldsymbol{t} = t(\cos\alpha, \sin\alpha, 0). \#(S10)$$

Then, the Berry curvature of the tilt Weyl nodes in Eq. (S8) and Eq. (S9) can be obtained, which can be utilized to solve Eq. (S7) for the distribution function and give the final form for the conductivity as follows:

$$\begin{aligned}
\sigma_{xx} = \frac{2e^2\tau}{(2\pi)^3\hbar^2}\Bigg[ & -4\pi v \frac{t - v\tanh^{-1}\frac{t}{v}}{\hbar t^3}\mu^2\cos^2\alpha + \pi v^2 \frac{2vt + 2(t^2 - v^2)\tanh^{-1}\frac{t}{v}}{\hbar t^3(v^2 - t^2)}\mu^2\sin^2\alpha - 4\pi etB\cos\alpha\cos\theta \\
& - \pi e \frac{-6t^5 + 10v^2 t^3 - 6v^4 t + 6v(t^2 - v^2)^2 \tanh^{-1}\frac{t}{v}}{3t^4} B\cos^2\alpha \cos(\theta - \alpha) \\
& + \pi e v^2 \frac{2t^3 - 3v^2 t + 3v(v^2 - t^2)\tanh^{-1}\frac{t}{v}}{3t^4} B\sin^2\alpha \cos(\theta - \alpha) \\
& - 2\pi e \frac{5v^2 t^3 - 3v^4 t + 3v(t^2 - v^2)^2 \tanh^{-1}\frac{t}{v}}{3t^4} B\sin\alpha\cos\alpha\sin(\theta - \alpha) + \pi e^2 \hbar \frac{t^2 v + v^3}{\mu^2} B^2 \cos^2\theta \\
& + \pi e^2 \hbar \frac{7t^2 v + v^3 + (4t^2 v + 2v^3)\cos^2(\theta - \alpha)}{15\mu^2} B^2 \cos^2\alpha + \pi e^2 \hbar \frac{v^3 + 2v^3 \sin^2(\theta - \alpha)}{15\mu^2} B^2 \sin^2\alpha \\
& - 4\pi e^2 \hbar \frac{t^2 v + v^3}{15\mu^2} B^2 \sin\alpha\cos\alpha\sin(\theta - \alpha)\cos(\theta - \alpha) \\
& - 2\pi e^2 \hbar \frac{13t^2 v + 5v^3}{15\mu^2} B^2 \cos\alpha\cos\theta\cos(\theta - \alpha) \\
& + 2\pi e^2 \hbar \frac{t^2 v + 5v^3}{15\mu^2} B^2 \sin\alpha\cos\theta\sin(\theta - \alpha) \Bigg], \quad (S11)
\end{aligned}$$

$$\sigma_{yx} = \frac{2e^2\tau}{(2\pi)^3\hbar^2}\Bigg[-4\pi v \frac{t - v\tanh^{-1}\frac{t}{v}}{\hbar t^3}\mu^2 \sin\alpha\cos\alpha - \pi v^2 \frac{2vt + 2(t^2 - v^2)\tanh^{-1}\frac{t}{v}}{\hbar t^3(v^2 - t^2)}\mu^2 \sin\alpha\cos\alpha - 2\pi etB\sin(\alpha + \theta)$$

$$-\pi e \frac{-6t^5 + 10v^2t^3 - 6v^4t + 6v(t^2 - v^2)^2\tanh^{-1}\frac{t}{v}}{3t^4} B\sin\alpha\cos\alpha\cos(\theta - \alpha)$$

$$-\pi e v^2 \frac{2t^3 - 3v^2t + 3v(v^2 - t^2)\tanh^{-1}\frac{t}{v}}{3t^4} B\sin\alpha\cos\alpha\cos(\theta - \alpha)$$

$$+\pi e \frac{5v^2t^3 - 3v^4t + 3v(t^2 - v^2)^2\tanh^{-1}\frac{t}{v}}{3t^4} B\cos 2\alpha\sin(\theta - \alpha) + \pi e^2\hbar\frac{t^2v + v^3}{\mu^2}B^2\sin\theta\cos\theta$$

$$+\pi e^2\hbar \frac{7t^2v + v^3 + (4t^2v + 2v^3)\cos^2(\theta - \alpha)}{15\hbar^2\mu^2}B^2\sin\alpha\cos\alpha$$

$$-\pi e^2\hbar \frac{v^3 + 2v^3\sin^2(\theta - \alpha)}{15\hbar^2\mu^2}B^2\sin\alpha\cos\alpha - 2\pi e^2\hbar\frac{t^2v + v^3}{15\hbar^2\mu^2}B^2\cos 2\alpha\sin(\theta - \alpha)\cos(\theta - \alpha)$$

$$-\pi e^2\hbar\frac{13t^2v + 5v^3}{15\hbar^2\mu^2}B^2\sin(\theta + \alpha)\cos(\theta - \alpha)$$

$$-\pi e^2\hbar\frac{t^2v + 5v^3}{15\hbar^2\mu^2}B^2\cos(\theta + \alpha)\sin(\theta - \alpha)\Bigg]. \quad (S12)$$

The Hall conductivity in Eq. (S12) can be utilized to fit the experimental data in Fig. 4f of the main text. The parameters in Fig. 4(f) are as follows: v = $4.6 \times 10^5$ m/s, t = $4.0 \times 10^5$ m/s, μ = 8 meV, τ = 0.4 ps, and α = $-75.5°$. Here, v is Fermi velocity, t is the tilt of Weyl cones, μ is chemical potential, τ is relaxation time and α labels the angle between tilt vector and current. The relaxation time and the Fermi velocity are in agreement with other experiments.

## II. Low energy effective model of ZrTe$_5$ in magnetic field

Previous DFT calculation found that 3D ZrTe$_5$ crystals are topological insulators [5]. Based on the DFT results, R. Y. Chen *et al.* constructed an effective model of ZrTe$_5$ with symmetry analysis [6]. The effective model consists of four basis states as shown in Eq.(S13). Here $\tau_z = \pm 1$ marks states of different orbitals, and σ's are Pauli matrices in spin space. Linear expansion around Γ point yields the following Dirac-like Hamiltonian, [6]

$$H_0 = m\tau_z + \hbar(v_x k_x t_x \sigma_z + v_y k_y \tau_y + v_z k_z \tau_x \sigma_x), \#(S13)$$

which describes the parent insulating state. Here $m$ is the Dirac mass, and $x, y, z$ axes above correspond to $a, c, b$ crystal axes, respectively.

In magnetic field, the Zeeman effect has to be considered, and there is an additional term in the Hamiltonian, [6]

$$H_{Zeeman} = -\frac{1}{2}\mu_B g \boldsymbol{\sigma} \cdot \boldsymbol{B},$$

where $\mu_B$ is Bohr magneton, and $g$ is the Landé g-factor. Further, we include some general quadratic dispersion around $\Gamma$ point,

$$H_1 = C_1 k_x^2 + C_2 k_y^2 + C_3 k_z^2.$$

For the complete Hamiltonian,

$$H = H_0 + H_1 + H_{Zeeman}, \#(S14)$$

the dispersion satisfies

$$E = C_1 k_x^2 + C_2 k_y^2 + C_3 k_z^2 \pm \sqrt{m^2 + \hbar^2(\boldsymbol{v}\cdot\boldsymbol{k})^2 + \frac{\mu_B^2 g^2 B^2}{4} \pm \hbar\mu_B g\sqrt{(B_z v_x k_x + B_x v_z k_z)^2 + B^2 v_y^2 k_y^2 + B^2 m^2}}, \#(S15)$$

where $\boldsymbol{v} = (v_x, v_y, v_z)$, and $\boldsymbol{k} = (k_x, k_y, k_z)$.

To incorporate with our experiment setup that the magnetic field lies in the $ac$-plane (the $xy$-plane here), we set $B_z = 0$. When the magnetic field is larger than the critical value $B_c = \left|\frac{2m}{\mu_B g}\right|$, the Hamiltonian above describes a time-reversal-symmetry breaking Weyl semimetal, with a pair of Weyl points at $\left(0, \pm\sqrt{\frac{\mu_B^2 g^2 B^2}{4} - m^2}, 0\right)$. The quadratic term $C_2 k_y^2$ in $H_1$ provides opposite additional velocities around the two Weyl points, which practically makes the two Weyl cones tilt along $y$-axis.

**Figures**

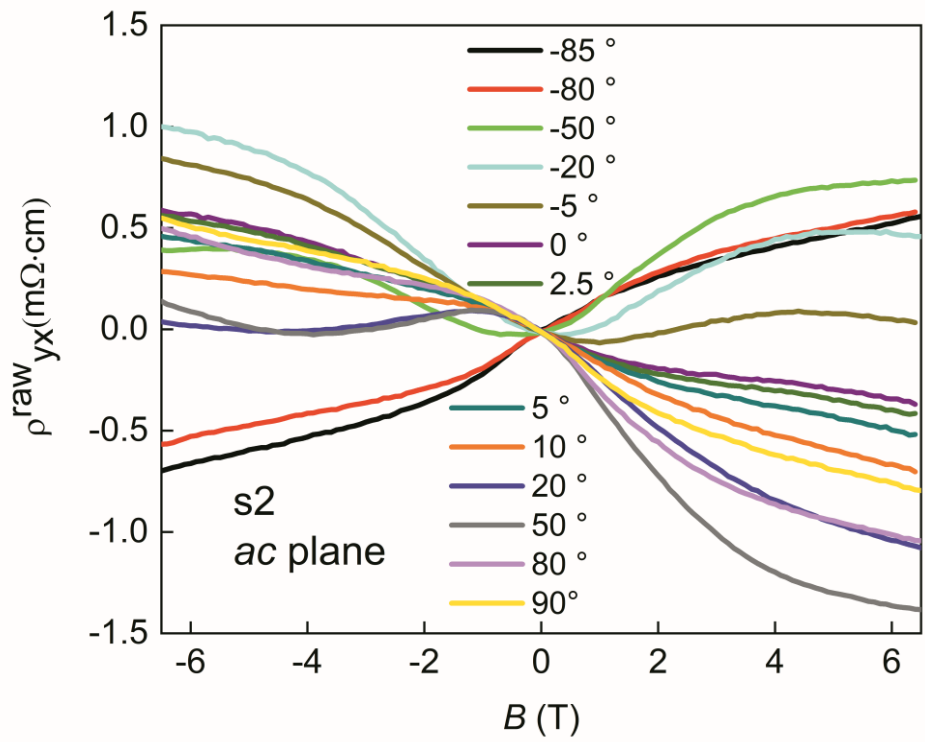

Fig. S1. Raw Hall data of ZrTe$_5$ device s2 detected at selected angles in PPMS.

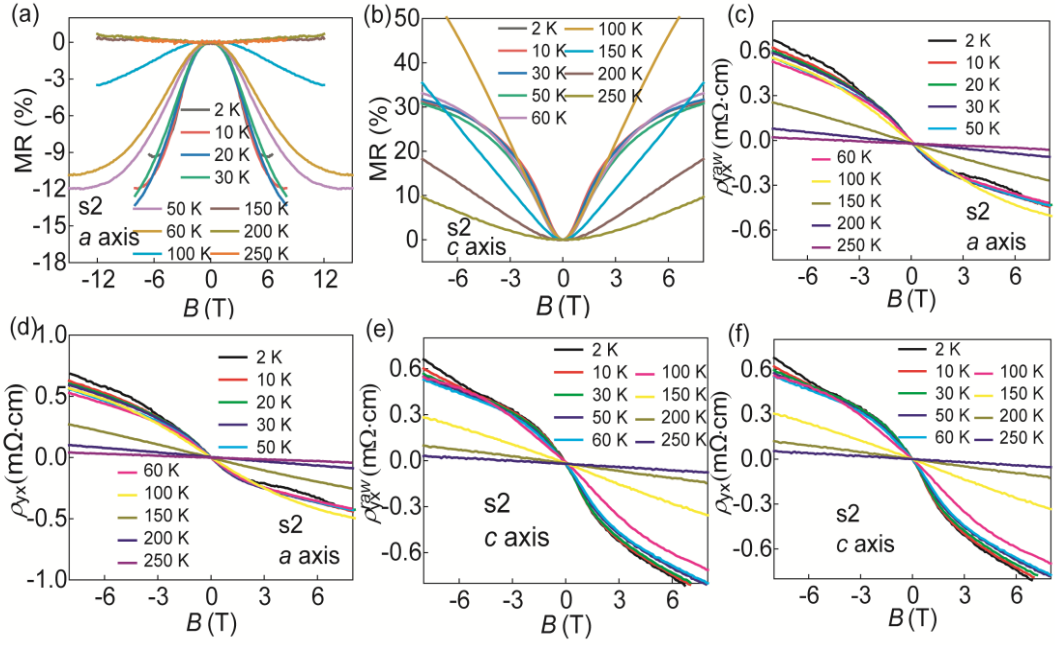

Fig. S2. Magnetoresistance ratio, Hall resistivity vs. B at $B \parallel I(\theta = 0°)$ and $B \perp I(\theta = 90°)$ at various temperatures of ZrTe$_5$ device s2. (a),(b) Magnetoresistance ratio vs B at *a* axis and *c* axis at various temperatures. (c),(d) Raw Hall data and data after subtracting $\rho_{xx}$ caused by electrodes misalignment at $B \parallel I$ at various temperatures. (e),(f) Raw Hall data and data after subtracting $\rho_{xx}$ caused by electrodes mismatch at $B \perp I$ at various temperatures.

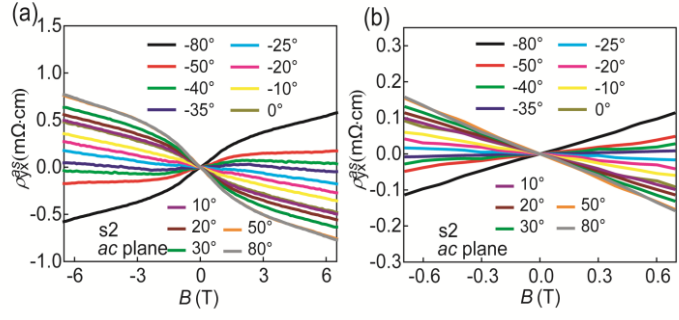

Fig. S3. Angular and temperature dependence of antisymmetric in-plane Hall resistivity of $ZrTe_5$ device s2 detected in PPMS. Magnetic field lies in *ac* plane. (a) Hall resistivity detected at selected angles from -6.5 T to 6.5 T. (b) Hall resistivity detected at selected angles from -0.7 T to 0.7 T.

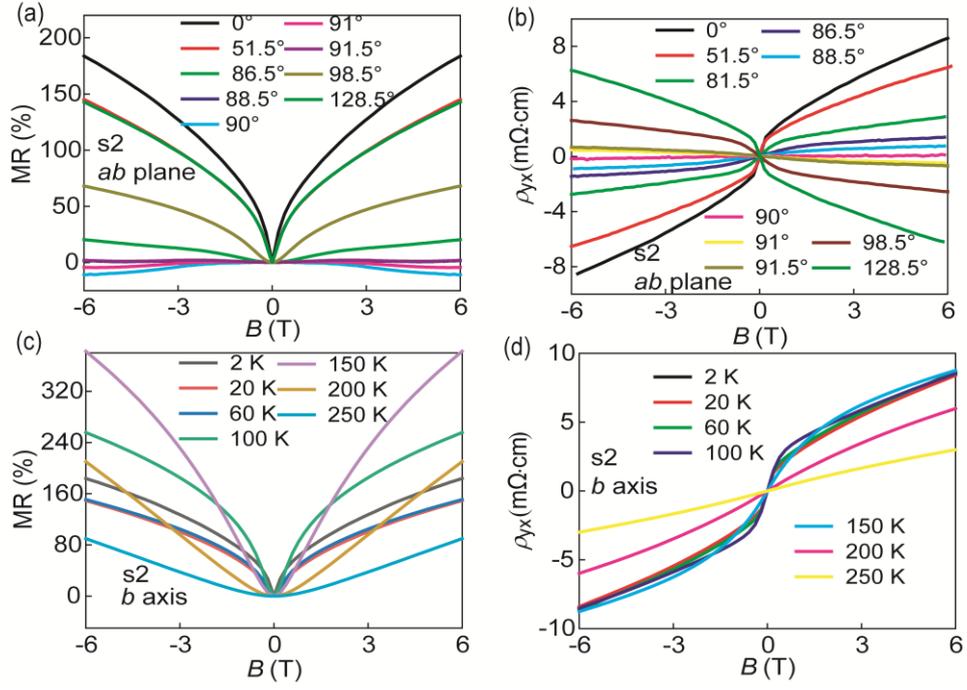

Fig. S4. Angular and temperature dependence of transport properties in *ab* plane of ZrTe$_5$ device s2. Magnetic field lies in *ab* plane. (a),(b) MR behavior and Hall resistivity at selected angles. Negative LMR is detected in a range of 3°. (c),(d) MR and Hall resistivity versus $B$ at various temperatures at $\boldsymbol{B} \parallel b$ axis.

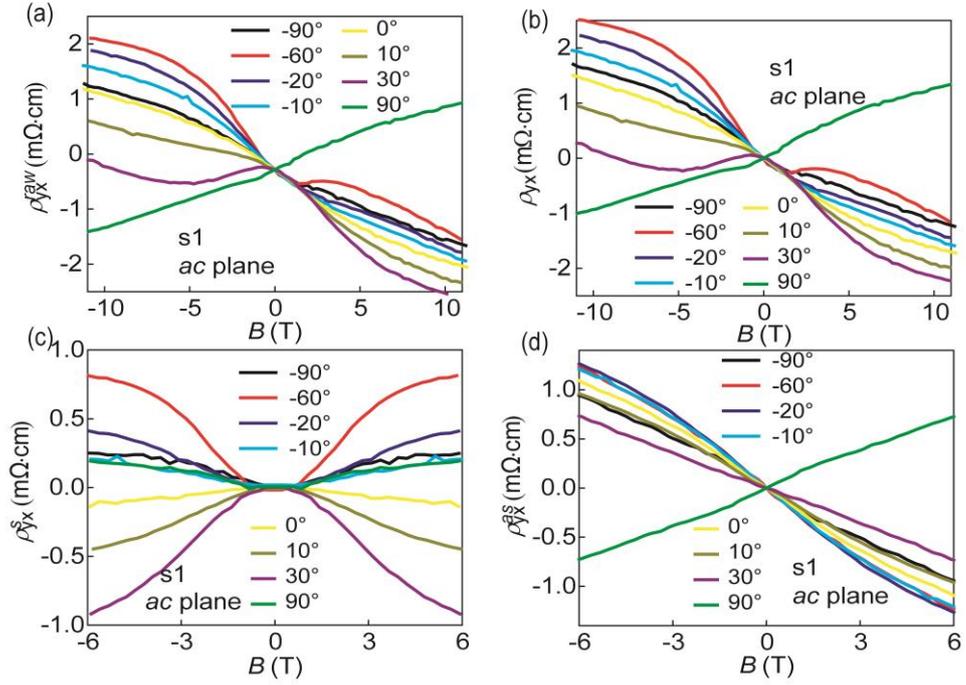

Fig. S5. Hall resistivity detected in ZrTe$_5$ device s1. (a) Raw Hall resistivity versus $B$ data at selected angles. (b) Hall resistivity at selected angles after subtracting $\rho_{xx}$ caused by electrodes mismatch. (c),(d) symmetric and antisymmetric in-plane Hall resistivity versus $B$ at selected angles.